\documentclass[]{spie}  
\usepackage[]{graphicx}

\title{Pauli graph and finite projective lines/geometries}

\author{Michel Planat\supit{a} and Metod Saniga\supit{b}
\skiplinehalf
\supit{a}Institut FEMTO-ST, CNRS, D\' epartement LPMO, 32 Avenue del'Observatoire,\\ F-25044 Besan\c con, France; \\
\supit{b}Astronomical Institute, Slovak Academy of Sciences, \\SK-05960 Tatransk\' a Lomnica, Slovak Republic}

\authorinfo{Further author information: (Send correspondence to Michel Planat)\\Michel Planat: E-mail: michel.planat@femto-st.fr, Telephone: +33 (0)3 81 85 39 57\\  Metod Saniga: E-mail: msaniga@astro.sk}

\begin{document}
  \maketitle

\begin{abstract}
The commutation relations between the generalized Pauli operators of $N$-qudits (i.\,e., $N$ $p$-level quantum systems), and the structure of their maximal sets of commuting bases, follow a nice graph theoretical/geometrical pattern. One may identify vertices/points with the operators so that edges/lines join commuting pairs of them to form the so-called Pauli graph $\mathcal{P}_{p^N}$. As per two-qubits ($p=2$, $N=2$) all basic properties and partitionings of this graph are embodied in the geometry of the symplectic generalized quadrangle of order two, $W(2)$.
The structure of the two-qutrit ($p=3$, $N=2$) graph is more involved; here it turns out more convenient to deal with its dual
in order to see all the parallels with the two-qubit case and its surmised relation with the geometry of generalized quadrangle $Q(4,3)$, the dual of $W(3)$.
Finally, the generalized adjacency graph for multiple ($N > 3$) qubits/qutrits is shown to follow from symplectic polar spaces of order two/three.  The relevance of these mathematical concepts to mutually unbiased bases and to quantum entanglement is also highlighted in some detail.

\end{abstract}



\section{INTRODUCTION}
\label{sect:intro}  
\noindent
The structure of commuting/non-commuting relations between $N$-qudit observables is at the heart of the peculiarities and strangeness of quantum mechanics. Its understanding is central to explain quantum complementarity, quantum entanglement, as well as other conceptual (or practical) issues like no-cloning, quantum teleportation, quantum cryptography and quantum computing, to mention a few.

In Hilbert spaces of prime-power dimension $d=p^N$, $p$ a prime, the $(d^2-1)$ generalized Pauli operators organize themselves into $(d+1)$ disjoint sets, each one containing a maximum set of $(d-1)$ mutually commuting operators. In this paper one deals with the Pauli graph $\mathcal{P}_d$ associated with an $N$-qudit system by regarding the operators as vertices, and joining any pair of commuting ones by an edge. The two-qubit system was studied in much detail in our previous papers\cite{Planat1,Planat2} and found to be related to a well-known finite geometry, namely the generalized quadrangle of order two (see Sec. 2 for details). In this paper, we also analyze the two-qutrit case and give important
clues that indicate that the geometry behind the corresponding Pauli graph and/or its dual is that of the symplectic
generalized quadrangle of order three, $W(3)$, and/or its dual, $Q(4,3)$.
Finally, we give some hints for generalizations to multiple qudits ($N>2$), employing symplectic polar spaces of order $N$ as the relevant finite geometries. For more details about the technicalities of the paper, the reader
is referred to consult Refs.~\citenum{Planat2} and \citenum{Saniga}--\citenum{Batten}.

Recently, we have found that an important class of finite geometries could be used for coordinatizing some subsets of the Pauli graphs (and, so, some subsets  of generalized quadrangles and/or polar spaces) --- projective lines defined over finite rings.\cite{Saniga1}$^-$\cite{Saniga2}  Given an associative ring $R$ with unity and $GL_{2}(R)$, the general linear group of invertible two-by-two matrices with entries in $R$, a pair $(\alpha,\beta)$ is called admissible over $R$ if there exist $\gamma,\delta \in R$ such that $\left(
\begin{array}{cc}
\alpha & \beta \\
\gamma & \delta \\
\end{array}
\right) \in GL_{2}(R)$. The projective line over $R$ is
defined as the set of equivalence classes of ordered pairs
$(\varrho \alpha, \varrho \beta)$, where $\varrho$ is a unit of
$R$ and $(\alpha, \beta)$ admissible \cite{bh1,Saniga2}. Such a
line carries two non-trivial, mutually complementary relations of
neighbor and distant. In particular, its two distinct points $X$:
$(\varrho \alpha, \varrho \beta)$ and $Y$: $(\varrho \gamma,
\varrho \delta)$ are called {\it neighbor} if $\left(
\begin{array}{cc}
\alpha & \beta \\
\gamma & \delta \\
\end{array}
\right) \notin GL_{2}(R)$ and {\it distant} otherwise. The
corresponding graph takes the points as vertices and its edges
link any two mutually neighbor points. For $R$ being the finite field $\mathbf{F}_k$, of order $k$, (the
graph of) the projective line lacks any edge, being an independent
set of cardinality $k+1$, or a $(k+1)$-coclique. Edges appear only
for a line over a ring featuring zero-divisors, and their number
is proportional to the number of zero-divisors and/or maximal
ideals of the ring concerned (see, e.\,g., Refs. \citenum{Saniga1}--\citenum{Saniga2}
for a comprehensive account of the structure of finite projective
ring lines). Projective lines of importance for our model will be the line defined over the
(non-commutative) ring of full $2\times 2$ matrices with
coefficients in $\mathcal{Z}_2$, as well as the lines defined over
three distinct types of rings of order four and characteristic two
\cite{Saniga}.

The paper is organized as follows: Secs.\,2 and 3 deal, respectively, with the Pauli graphs and the associated geometry of two-qubit and two-qutrit
systems and Sec.\,4 highlights important geometric-combinatorial clues about arbitrary $N$-qudits.

\section{The Pauli graph of two-qubits}
\label{twoqubits}
\noindent
Let us consider the fifteen tensor products $\sigma_i \otimes \sigma_j$, $i,j \in \{1,2,3,4\}$ and $(i,j)\neq (1,1)$, of Pauli matrices $\sigma_i= (I_2,\sigma_x, \sigma_y,\sigma_z)$, where  $I_2=\left(\begin{array}{cc}1 & 0 \\0 & 1\\\end{array}\right)$, $\sigma_x=\left(\begin{array}{cc}0 & 1 \\1 & 0\\\end{array}\right)$, $\sigma_z=\left(\begin{array}{cc}1 & 0 \\0 & -1\\\end{array}\right)$ and $\sigma_y=i \sigma_x \sigma_z$ and label them as follows $1=I_2 \otimes \sigma_x$, $2=I_2 \otimes \sigma_y$, $3=I_2 \otimes \sigma_z$, $a=\sigma_x \otimes I_2$, $4=\sigma_x \otimes \sigma_x$\ldots, $b=\sigma_y \otimes I_2$,\ldots , $c=\sigma_z \otimes I_2$,\ldots  Joining two distinct mutually commuting operators by an edge, one obtains the Pauli graph $\mathcal{P}_4$ with incidence matrix as shown in Table 1. The main invariants of $\mathcal{P}_4$ and those of some of its most important subgraphs are listed in Table 2.
As it readily follows from Table 1, $\mathcal{P}_4$ is $6$-regular and, so, intricately connected with the complete graphs $K_n$, $n=5$, $6$ or $7$. First, one checks that $\mathcal{P}_4 \cong \hat{L}(K_6)$, i.\,e., it is isomorphic to the complement of line graph of $K_6$.  Next, computing its minimum vertex cover (Table 2), one recovers the Petersen graph $PG \equiv \hat{L}(K_5)$. Finally, $\mathcal{P}_4$ is also found to be isomorphic to the minimum vertex cover of $\hat{L}(K_7)$. Now, we turn to remarkable partitionings/factorizations and the corresponding distinguished subgraphs of $\mathcal{P}_4$.

%
\begin{table}[h]
\begin{center}
\begin{tabular}{|r|r|r|r|r|r|r|}
\hline
$O_4$&1& $I_4$ &1& $I_4$ &1& $I_4$ \\
\hline
1&0&1&0&0&0&0\\
\hline
$I_4$&1& $O_4$ &0& $\hat{I}_4$ &0& $\hat{I}_4$ \\
\hline
1&0&0&0&1&0&0\\
\hline
$I_4$&0& $\hat{I}_4$ &1& $O_4$ &0& $\hat{I}_4$ \\
\hline
1&0&0&0&0&0&1\\
\hline
$I_4$&0& $\hat{I}_4$&0 & $\hat{I}_4$ &1& $O_4$ \\
\hline
\end{tabular}
\caption{Structure of the incidence matrix of the two-qubit Pauli graph $\mathcal{P}_4$. The single rows/columns at the reference points $a$, $b$ and $c$ separate the matrix blocks $O_4$, $I_4$ and $\hat{I}_4$. }
\end{center}
\label{graphP4}
\end{table}
\begin{table}[htb]
\begin{center}
\begin{tabular}{|r|r|r||r|r||r|r|}
\hline
 $G$&$\mathcal{P}_4$&$PG \cong MVC$&$MS$&$BP$&$FP$&$CB$\\
 \hline
$v$ &$15$&$10$&$9$&$6$&$7$&$8$\\
 $e$&$45$&15&$18$&$9$&$9$&$12$\\
 $_{spec(G)}$&$_{\{-3^5,1^9,6\}}$&$_{\{-2^4,1^5,3\}}$&$_{\{-2^4,1^4,4\}}$&$_{\{-3,0^4,3\}}$&$_{\{-2,-1^3,1^2,3\}}$&$_{\{-3,-1^3,1^3,3\}}$\\
 $g(G)$&$3$&5&$3$&$4$&$3$&$3$\\
 $\kappa(G)$&$4$&3&$3$&$2$&$3$&$2$\\
 \hline
\end{tabular}
\end{center}
\label{invariantsP22}
\caption{The main invariants of the Pauli graph $\mathcal{P}_4$ and its subgraphs, including its
minimum vertex covering $MVC$ isomorphic to the Petersen graph $PG$. For the remaining symbols, see Refs.\,\citenum{Planat2} and \citenum{Harary}.}
\end{table}
\begin{figure}[h]
\centerline{\includegraphics[width=9.0truecm,clip=]{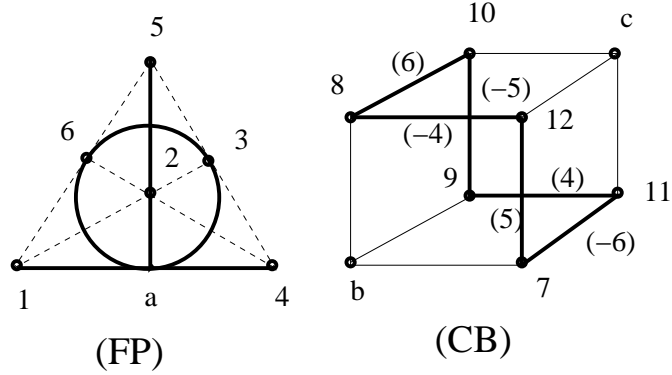}}
\caption{Partitioning of $\mathcal{P}_4$ into a pencil of lines
in the Fano plane ($FP$) and a cube ($CB$). In $FP$ any two
observables on a line map to the third one on the same line. In
$CB$ two vertices joined by an edge map to points/vertices in
$FP$. The map is explicitly given for the entangled Hamiltonian
path by labels on the corresponding edges.}
\end{figure}
\begin{figure}[h]
\centerline{\includegraphics[width=9.0truecm,clip=]{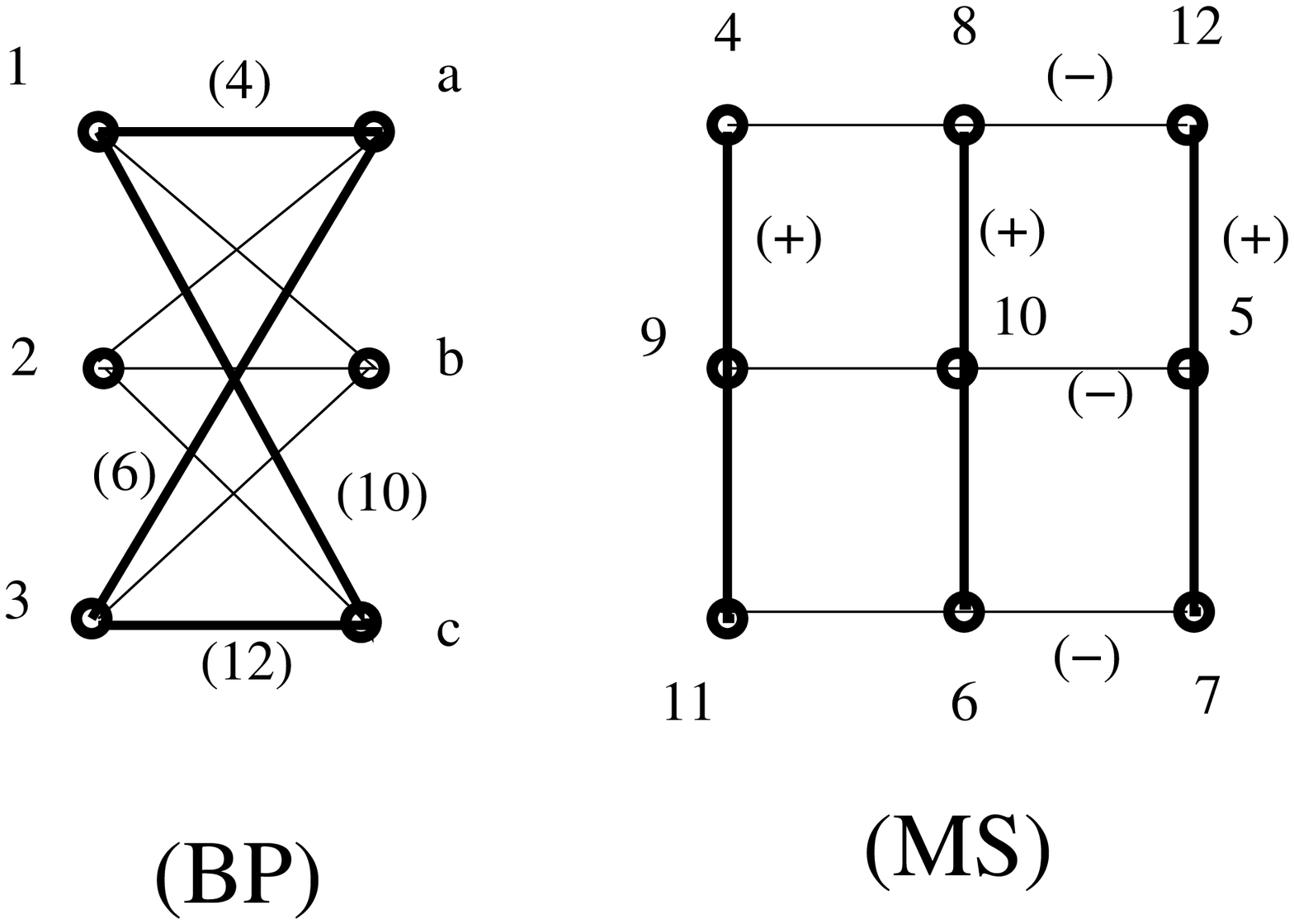}}
\caption{Partitioning of $\mathcal{P}_4$ into an unentangled bipartite graph ($BP$) and a fully entangled Mermin square ($MS$). In $BP$ two vertices on any edge map to a point in $MS$ (see the labels of the edges on a selected Hamiltonian path). In $MS$ any two vertices on a line map to the third one. Operators on all six lines carry a base of entangled states. The graph is polarized, i.e., the product of three observables in a row is $-I_4$, while in a column it is $+I_4$.}
\end{figure}
\subsection{The ``Fano pencil" $FP$ and the cube $CB$}
\noindent
We shall first tackle the 7+8 partitioning of the graph which can, for example, be realized by the following subgraphs/subsets: $FP = \langle 1,2,3,a,4,5,6 \rangle$ and $CB = \langle b,7,8,9,c,10,11,12 \rangle$ ---
see Fig.\,1. The subgraph $FP$ can also be regarded as a line pencil in the Fano plane \cite{Planat1,Polster} as well as a hyperplane of $W(2)$ \cite{Saniga}; the number of choices for this partitioning is obviously equal to the number of the vertices of the full graph (see Ref.~\citenum{Planat1} for another choice). It is easy to observe that two vertices on one line of $FP$ map to the third one on the same line, i.e., $1.a=4$, $2.a=5$ and $3.a=6$. The three observables are found to share a common base of $4$-dimensional vectors;  for this particular choice, the lines in the Fano pencil $FP$ feature unentangled 2-qubit bases. In addition, an edge of $CB$ is mapped to a vertex of $FP$, e.\,g., $8.10=6$, $8.12=-4$, etc. In particular there is an Hamiltonian cycle of length $6$ (shown with thick lines) in the cube graph $CB$ which features six bases of entangled states. It is worth mentioning here that  in Ref.~\citenum{Planat1} the projective lines over direct product of rings of the type $\mathcal{Z}_2^{\times n}$, $n=2,3,4$, were used to tackle this kind of partitioning. With these lines it was possible to grasp the structure of the two subsets, but not the coupling between them; to get a complete picture required employing a more abstract projective line with a more involved structure \cite{Saniga}.

\subsection{The Mermin square $MS$ and the bipartite part $BP$}
\noindent
We shall focus next on the 9+6 partitioning which can be illustrated, for example, by the subgraphs $BP=\langle 1,2,3,a,b,c\rangle$ and $MS=\langle 4,5,6,7,8,9,10,11,12\rangle$ --- see Fig.\,2. The $BP$ part is easily recognized as the bipartite graph $K[3,3]$, while the $MS$ part is a $4$-regular graph.  There is a map from the edges of $BP$ to the vertices of $MS$, and a map from two vertices of a line in $MS$ to the third vertex on the same line. The bases defined by two commuting operators in $BP$ are unentangled. By contrast, operators on any row/column of $MS$ define an entangled base. A square/grid like the $MS$ was used by Mermin \cite{Mermin} --- and frequently referred to as a Mermin's square since then --- to provide a simple proof of the Kochen-Specker theorem in four dimensions. The proof goes as follows. One observes that the square is polarized in the sense that the product of three operators on any column equals $+ I_4$ (the $4 \times 4$ identity matrix), while the product of three observables on any row equals  $-I_4$. By multiplying all columns and rows one gets $-I_4$. This is, however, not the case for the eigenvalues of the observables; they all equal $\pm 1$ and their corresponding products always yield $+1$ because each of them appears in the product twice; once as the eigenvalue in a column and once as the eigenvalue in a row. The algebraic structure of mutually commuting operators thus contradicts that of their eigenvalues, which furnishes a proof of the Kochen-Specker theorem.
\begin{figure}[h]
\centerline{\includegraphics[width=9truecm,clip=]{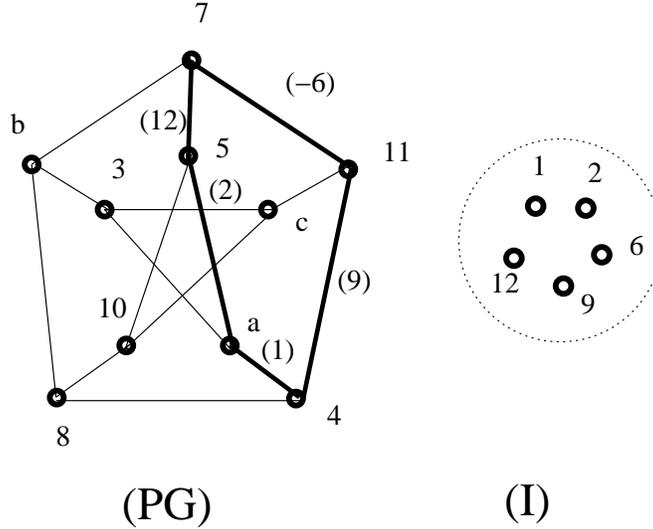}}
\caption{The partitioning of $\mathcal{P}_4$ into an independent set ($I$) and  the Petersen graph ($PG$), {\it aka} its minimum vertex cover. The two vertices on an edge of $PG$ correspond/map to a vertex in $I$ (as illustrated by the labels on the edges of a selected Hamiltonian path).}
\end{figure}
The $MS$ set is also recognized as a $(9_2,6_3)$ configuration for
any point is incident with two lines and any line is incident with
three points and does not change its shape if we reverse our
notation, i.\,e., join by an edge two mutually non-commuting
observables; in graph theoretical terms this means that the $MS$
equals its complement. Last but not least,
it needs to be mentioned that the $MS$ configuration represents also
the structure of the projective line over the product ring
$\mathcal{Z}_2 \times \mathcal{Z}_2$ if we identify the points
sets of the two and regard edges as joins of mutually {\it distant}
points; it was precisely this fact that
motivated our in-depth study of projective ring lines
\cite{Saniga1,Saniga2} and finally led to the discovery of the
relevant geometries behind two- and multiple-qubit systems
\cite{Planat1,Saniga,Saniga4}.
\begin{figure}[h]
\centerline{\includegraphics[width=7.0truecm,clip=]{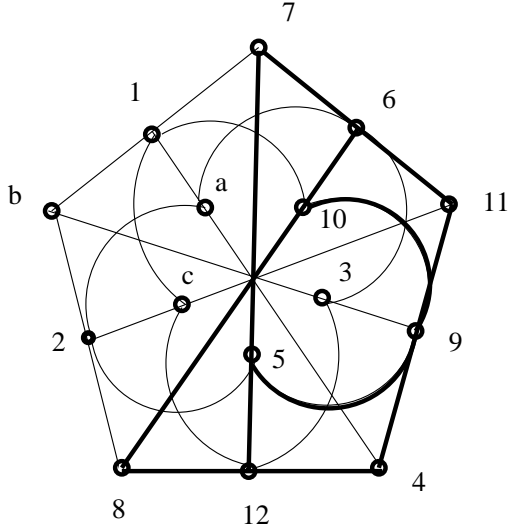}}
\caption{$W(2)$ as the {\it unique} underlying geometry of two-qubit systems. The Pauli operators correspond to the points and maximally commuting subsets of them to the lines of the quadrangle.  Three operators on each line have a common base; six out of fifteen such bases are entangled (the corresponding lines being indicated by boldfacing). }
\end{figure}
\subsection{The Petersen graph $PG$ and the maximum independent set $I$}
\noindent
The third fundamental partitioning of $\mathcal{P}_4$ comprises a maximum independent set $I$ and the Petersen graph $PG$ \cite{Saniga}. This can be done in six different ways, one of them featuring $I=\langle 1,2,6,9,12\rangle$ and  $PG=\langle 3,a,4,5,b,7,8,c,10,11\rangle$ --- see Fig.\,3.
As in the case of their cousins $CB$ and $BP$, the Petersen graph $PG$ admits a map of its edges into the vertices of the independent set $I$.

\subsection{Finite projective algebraic geometry underlying $\mathcal{P}_4$}
\subsubsection{$\mathcal{P}_4$ as the generalized quadrangle of order two --- $W(2)$}
\noindent
At this point we have dissected $\mathcal{P}_4$ to such an extent that we are ready to show the unique finite projective geometry hidden behind --- namely the {\it generalized quadrangle of order two}, $W(2)$ \cite{Saniga}. As already mentioned, $W(2)$ is the simplest thick generalized quadrangle endowed with fifteen points and the same number of lines, where every line features three points and, dually, every point is incident with three lines, and where every point is joined by a line (or, simply, collinear) with other six points \cite{Payne,Polster}. These properties can easily be grasped from the drawing of this object, dubbed for obvious reasons the doily, depicted in Fig.\,4; here, all the points are drawn as small circles, while lines are represented either by line segments (ten of them), or as segments of circles (the remaining five of them). To recognize in this picture $\mathcal{P}_4$ one just needs to identify the fifteen points of $W(2)$ with our fifteen generalized Pauli operators as explicitly illustrated, with the understanding that {\it collinear} means {\it commuting} (and, so, {\it non}-collinear reads {\it non}-commuting); the fifteen lines of $W(2)$ thus stand for nothing but fifteen {\it maximum} subsets of three mutually commuting operators each.

That $W(2)$ is indeed the right projective setting for
$\mathcal{P}_4$ stems also from the fact that it gives a nice
geometric justification for all the three basic
partitionings/factorizations of $\mathcal{P}_4$. To see this, we
just employ the fact that $W(2)$ features three distinct kinds of
geometric hyperplanes \cite{Payne}: 1) a {\it perp}-set
($H_{cl}(X)$), i.\,e., a set of points collinear with a given
point $X$, the point itself inclusive (there are 15 such
hyperplanes);  2) a {\it grid} ($H_{gr}$) of nine points on six
lines, {\it aka} a slim generalized quadrangle of order $(2,1)$
(there are 10 such hyperplanes); and 3) an {\it ovoid} ($H_{ov}$),
i.\,e., a set of (five) points that has exactly one point in
common with every line (there are six such hyperplanes). One then
immediately sees \cite{Saniga} that a perp-set is identical with a
Fano pencil, a grid answers to a Mermin square and, finally, an
ovoid corresponds to a maximum independent set. Because of
self-duality of $W(2)$, each of the above introduced hyperplanes
has its dual, line-set counterpart. The most interesting of them
is the dual of an ovoid, usually called a {\it spread},  i.\,e., a
set of (five) pairwise disjoint lines that partition the point
set; each of six different spreads of $W(2)$ represents such a
pentad of mutually disjoint maximally commuting subsets of
operators whose associated bases are {\it mutually unbiased}
\cite{Planat1,Lawrence1}. It is also important to mention a {\it
dual grid}, i.\,e., a slim generalized quadrangle of order
$(1,2)$, having a property that the three operators on any of its
nine lines share a base of {\it un}entangled states. It is
straightforward to verify that these lines are defined by the
edges of a $BP$; each of the remaining six lines (fully located in
the corresponding/complementary $MS$) carries a base of {\it
entangled} states (see Fig.\,4).

\subsubsection{$\mathcal{P}_4$ and the projective line over the full two-by-two matrix ring over $\mathcal{Z}_2$}
\label{ringline}
$W(2)$ is found as a {\it sub}geometry of many interesting projective configurations and spaces \cite{Payne,Polster}. We will now briefly examine a couple of such embeddings of $W(2)$ in order to reveal further intricacies of its structure and, so, to get further insights into the structure of the two-qubit Pauli graph.

We shall first consider an embedding of $W(2)$ in the projective line defined over the ring $\mathcal{Z}_2^{2 \times 2}$ of full $2 \times 2$ matrices with $\mathcal{Z}_2$-valued coefficients,
\begin{equation}
 \mathcal{Z}_2^{2 \times 2}    \equiv \left\{ \left(
\begin{array}{cc}
\alpha & \beta \\
\gamma & \delta \\
\end{array}
\right) \mid ~ \alpha, \beta, \gamma, \delta \in \mathcal{Z}_2 \right\},
\end{equation}
because it was this projective ring geometrical setting where the relevance of
the structure $W(2)$ for two-qubits was discovered \cite{Saniga}.
To facilitate our reasonings, we label the matrices of $\mathcal{Z}_2^{2 \times 2}$ in the following way
\footnotesize
\begin{eqnarray}
\footnotesize
&&~1' \equiv \left(
\begin{array}{cc}
1 & 0 \\
0 & 1 \\
\end{array}
\right),~2' \equiv \left(
\begin{array}{cc}
0 & 1 \\
1 & 0 \\
\end{array}
\right),~ 3' \equiv \left(
\begin{array}{cc}
1 & 1 \\
1 & 1 \\
\end{array}
\right),~
4' \equiv \left(
\begin{array}{cc}
0 & 0 \\
1 & 1 \\
\end{array}
\right),
~5' \equiv \left(
\begin{array}{cc}
1 & 0 \\
1 & 0 \\
\end{array}
\right),~6' \equiv \left(
\begin{array}{cc}
0 & 1 \\
0 & 1 \\
\end{array}
\right),
\nonumber \\&&~ 7' \equiv \left(
\begin{array}{cc}
1 & 1 \\
0 & 0 \\
\end{array}
\right),
8' \equiv \left(
\begin{array}{cc}
0 & 1 \\
0 & 0 \\
\end{array}
\right),~9' \equiv \left(
\begin{array}{cc}
1 & 1 \\
0 & 1 \\
\end{array}
\right),~10' \equiv \left(
\begin{array}{cc}
0 & 0 \\
1 & 0 \\
\end{array}
\right),~11' \equiv \left(
\begin{array}{cc}
1 & 0 \\
1 & 1 \\
\end{array}
\right),~12' \equiv \left(
\begin{array}{cc}
0 & 1 \\
1 & 1 \\
\end{array}
\right),
\nonumber \\&&
13' \equiv \left(
\begin{array}{cc}
1 & 1 \\
1 & 0 \\
\end{array}
\right),~14' \equiv \left(
\begin{array}{cc}
0 & 0 \\
0 & 1 \\
\end{array}
\right),~15' \equiv \left(
\begin{array}{cc}
1 & 0 \\
0 & 0 \\
\end{array}
\right),~0' \equiv \left(
\begin{array}{cc}
0 & 0 \\
0 & 0 \\
\end{array}
\right),
\normalsize
\end{eqnarray}
\normalsize
and see that $\{1',2',9',11',12',13'\}$ are units (i.\,e., invertible matrices) and   $\{0',3',4',5',6',7',8',10',14',15'\}$ are zero-divisors (i.\,e., matrices with vanishing determinants), with 0' and 1' being, respectively, the additive and multiplicative identities of the ring.  Employing the definition of a projective ring line given in Refs.~\citenum{Saniga1} and \citenum{bh1}, it
is a routine, though a bit cumbersome, task to find out that the line over  $\mathcal{Z}_2^{2 \times 2}$ is endowed with 35 points whose coordinates, up to left-proportionality by a unit, read as follows
\begin{eqnarray}
&&(1',1'),~(1',2'),~(1',9'),~(1',11'),~(1',12'), (1',13'), \nonumber \\
&&(1',0'),~(1',3'),~(1',4'),~(1',5'),~(1',6'),~(1',7'),~(1',8'),~(1',10'),~(1',14'),~(1',15'), \nonumber \\
&&(0',1'),~(3',1'),~(4',1'),~(5',1'),~(6',1'),~(7',1'),~(8',1'),~(10',1'),~(14',1'),~(15',1'), \nonumber \\
&&(3',4'),~(3',10'),~(3',14'),~(5',4'),~(5',10'),~(5',14'),~(6',4'),~(6',10'),~(6',14').
\end{eqnarray}
Next, we pick up two mutually distant points of the line. Given the fact that
$GL_{2}(R)$ act transitively on triples of pairwise distant points \cite{bh1}, the two points can, without any loss of generality, be taken to
be the points $U_{0}:=(1,0)$ and $V_{0}:=(0,1)$. The points of $W(2)$ are then those points of the line which are either simultaneously distant
or simultaneously neighbor to $U_{0}$ and $V_{0}$. The shared distant points are, in this particular representation, (all the) six points
whose both entries are units,
\begin{eqnarray}
&&(1',1'),~(1',2'),~(1',9'), \nonumber \\
&&(1',11'),~(1',12'),~(1',13'),
\end{eqnarray}
whereas the common neighbors comprise (all the) nine points with both coordinates being zero-divisors,
\begin{eqnarray}
&&(3',4'),~(3',10'),~(3',14'), \nonumber \\
&&(5',4'),~(5',10'),~(5',14'),  \nonumber \\
&&(6',4'),~(6',10'),~(6',14'),
\end{eqnarray}
the two sets thus readily providing a ring geometrical explanation for a $BP+MS$ factorization of the algebra of the two-qubit Pauli operators, Fig.\,5, after the concept of mutually {\it neighbor} is made synonymous with that of mutually {\it commuting} \cite{Saniga}.
To see all the three factorizations within this setting it suffices to notice that the ring $\mathcal{Z}_2^{2 \times 2}$ contains as subrings all the {\it three} distinct kinds of rings of order four and characteristic two, viz. the (Galois) field $\mathbf{F}_4$, the local ring
$\mathcal{Z}_2[x]/\langle x^2 \rangle$, and the direct product ring $\mathcal{Z}_2 \times \mathcal{Z}_2$ \cite{mcd}, and check that the corresponding lines can be identified with the three kinds of geometric hyperplanes of $W(2)$ as shown in Table\,3 \cite{Saniga}.
\begin{table}[ht]
\begin{center}
\caption{Three kinds of the distinguished subsets of the
generalized Pauli operators of two-qubits ($\mathcal{P}_4)$) viewed either as the geometric
hyperplanes in the generalized quadrangle of order two ($W(2)$) or as
the projective lines over the rings of order four and
characteristic two residing in  the projective line over $\mathcal{Z}_2^{2 \times 2}$.}
\vspace*{0.4cm}
\begin{tabular}{llll}
\hline \hline
\vspace*{-.3cm} \\
$\mathcal{P}_4$ & set of five mutually    & set of six operators  & nine operators of a \\
   & non-commuting operators  & commuting with a given one & Mermin's square\\
   $W(2)$ & ovoid  &   perp-set$\setminus$\{reference point\} &   grid \\
Proj. Lines over & $\mathbf{F}_4 \cong \mathcal{Z}_2[x]/\langle x^2 + x + 1 \rangle$  &      $\mathcal{Z}_2[x]/\langle x^2 \rangle$  &   $\mathcal{Z}_2 \times \mathcal{Z}_2 \cong \mathcal{Z}_2[x]/\langle x(x+1) \rangle$  \\
\vspace*{-.3cm} \\
\hline \hline
\end{tabular}
\end{center}
\end{table}
\begin{figure}[htb]
\centerline{\includegraphics[width=11.3truecm,clip=]{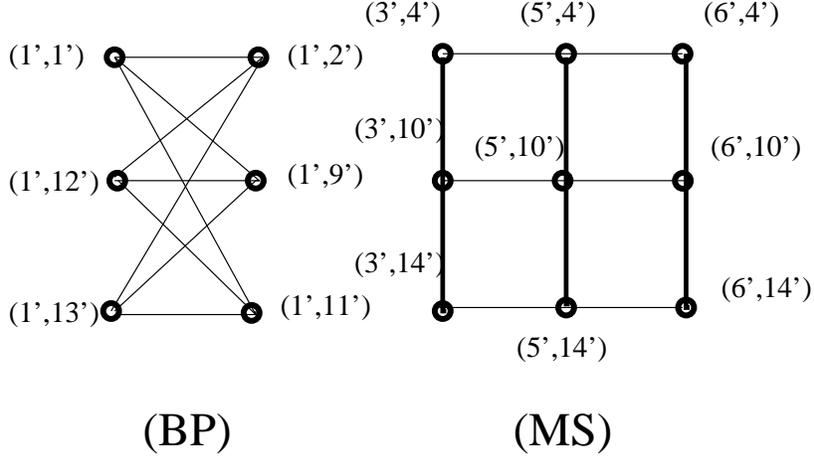}}
\caption{A $BP+MS$ factorization of $\mathcal{P}_4)$ in terms of the points of the subconfiguration of
the projective line over the full matrix ring $\mathcal{Z}_2^{2 \times 2}$; the points of the $BP$ have both coordinates units, whilst those of the $MS$ feature in both entries zero-divisors. The ``polarization" of the Mermin square is in this particular ring geometrical setting expressed by the fact that each column/row is characterized by the fixed value of the the first/second coordinate. Compare with Fig.\,2.}
\end{figure}

The other embedding of $W(2)$ is the one into the simplest projective space, $PG(3,2)$, as illustrated in Ref.~\citenum{Planat2}. This embedding is, in fact, a very close ally of the previous one due to a remarkable bijective correspondence between the points of the line over $\mathcal{Z}_2^{2 \times 2}$ and the lines of $PG(3,2)$ \cite{Thas}. $W(2)$ and $PG(3,2)$ are identical as the point sets,
whilst the fifteen lines of $W(2)$ are so-called totally isotropic lines with respect to a symplectic polarity of $PG(3,2)$ (see Sec.~\ref{polar}).

\section{The Pauli graph of two-qutrits}
\label{twoqutrits}
A complete orthonormal set of operators of a single-qutrit Hilbert space is \cite{Lawrence2}
\begin{equation}
\sigma_I=\{ I_3,Z,X,Y,V,Z^2,X^2,Y^2,V^2\},~~I=1,2,3,\dots,9,
\end{equation}
where $I_3$ is the $3 \times 3$ unit matrix, $Z=\left(\begin{array}{ccc}1 & 0&0 \\0 & \omega&0\\0&0& \omega^2\\\end{array}\right)$, $X=\left(\begin{array}{ccc}0 & 0&1 \\1 & 0&0\\0&1& 0\\\end{array}\right)$, $Y=XZ$, $V=X Z^2$ and $\omega=\exp\left(2 i \pi/3\right)$.
Labelling the two-qutrit Pauli operators as follows $1=I_3 \otimes \sigma_1$, $2=I_3 \otimes \sigma_2$, $\cdots$, $8=I_3 \otimes \sigma_8$, $a=\sigma_1 \otimes I_3$, $9=\sigma_1 \otimes \sigma_1$,\ldots, $b=\sigma_2 \otimes I_3$, $17=\sigma_2 \otimes \sigma_1$,\ldots , $c=\sigma_3 \otimes I_3$,$\ldots$, $h=\sigma_8 \otimes I_2$,$\ldots$, $72=\sigma_8 \otimes \sigma_8$, one obtains the incidence matrix of the two-qutrit Pauli graph $\mathcal{P}_9$ as shown in Table 4. Here, $B_8=\left(\begin{array}{cc}U & \hat{U} \\\hat{U} & U\\\end{array}\right)$ with $U=\left(\begin{array}{cccc}0 &0&0&0 \\1 & 0&1&0\\
1&0&0&1\\1&1&0&0\end{array}\right)$, $E_8=\left(\begin{array}{cc}0_4 & I_4 \\I_4 & 0_4\\\end{array}\right)$, $F_8=\left(\begin{array}{cc}I_4 & I_4 \\I_4 & I_4\\\end{array}\right)$, and $0_4$ and $I_4$ are the $4 \times 4$-dimensional all-zero and unit matrix, respectively.

\begin{table}[ht]
\begin{center}
\begin{tabular}{|r|r|r|r|r|r|r|r|r|r|r|r|r|r|r|r|r|}
\hline
$E_8$& 1&$F_8$&1 & $F_8$&1 & $F_8$&1 &$F_8$&1& $F_8$&1 & $F_8$&1 & $F_8$&1 &$F_8$\\
\hline
1&0&1&0&0&0&0&0&0&1&1&0&0&0&0&0&0\\
\hline
$F_8$& 1&$E_8$&0 & $B_8$&0 & $B_8$&0 & $B_8$&1 & $F_8$&0 & $\hat{B}_8$&0 & $\hat{B}_8$&0& $\hat{B}_8$\\
\hline
1&0&0&0&1&0&0&0&0&0&0&1&1&0&0&0&0\\
\hline
$F_8$&0 &$\hat{B}_8$&1 & $E_8$&0 & $\hat{B}_8$&0 & $B_8$&0 & $B_8$&1 & $F_8$&0 & $B_8$&0 &$\hat{B}_8$ \\
\hline
1&0&0&0&0&0&1&0&0&0&0&0&0&1&1&0&0\\
\hline
$F_8$&0 &$\hat{B}_8$&0 & $B_8$&1 & $E_8$&0 &$\hat{B}_8$&0 &$B_8$&0 & $\hat{B}_8$&1 &$F_8$&0 &$B_8$\\
\hline
1&0&0&0&0&0&0&0&1&0&0&0&0&0&0&1&1\\
\hline
$F_8$ &0& $\hat{B}_8$&0 &$\hat{B}_8$&0 & $B_8$&1 & $E_8$&0 & $B_8$&0 & $B_8$&0 & $\hat{B}_8$&1 & $F_8$\\
\hline
1&1&1&0&0&0&0&0&0&0&1&0&0&0&0&0&0\\
\hline
$F_8$&1 & $F_8$&0 & $\hat{B}_8$&0& $\hat{B}_8$&0& $\hat{B}_8$&1 & $E_8$&0 & $B_8$&0 & $B_8$&0 & $B_8$\\
\hline
1&0&0&1&1&0&0&0&0&0&0&0&1&0&0&0&0\\
\hline
$F_8$&0 & $B_8$&1 & $F_8$&0& $B_8$&0& $\hat{B}_8$&0 & $\hat{B}_8$&1 & $E_8$&0 & $\hat{B}_8$&0 & $B_8$\\
\hline
1&0&0&0&0&1&1&0&0&0&0&0&0&0&1&0&0\\
\hline
$F_8$&0 & $B_8$&0 & $\hat{B}_8$&1& $F_8$&0& $B_8$&0 & $\hat{B}_8$&0 & $B_8$&1 & $E_8$&0 & $\hat{B}_8$\\
\hline
1&0&0&0&0&0&0&1&1&0&0&0&0&1&1&0&1\\
\hline
$F_8$&0 & $B_8$&0 & $B_8$&0& $\hat{B}_8$&1& $F_8$&0 & $\hat{B}_8$&0 & $\hat{B}_8$&0 & $B_8$&1 & $E_8$\\
\hline
\end{tabular}
\label{twoqutrits}
\caption{Structure of the incidence matrix of the two-qutrit Pauli graph $\mathcal{P}_9$ }
\end{center}
\end{table}
\noindent
Computing the spectrum $\{-7^{15},-1^{40},5^{24},25\}$ one observes that the graph is regular, of degree $25$, but not strongly regular. The structure of observables in $\mathcal{P}_9$ is much more involved than in the case of two-qubits although it is still possible to recognize identifiable regular subgraphs.
In order to get necessary hints for the geometry behind this system, it necessitates
to pass to its dual graph, $\mathcal{W}_9$, i.\,e., the graph whose vertices are maximally commuting subsets
(MCSs) of  $\mathcal{P}_9$. To this end, let us first give a complete list of the latter:

\footnotesize
\begin{eqnarray}
&L_1=\{1,5,a,9,13,e,41,45\},~~L_2=\{2,6,a,10,14,e,42,46\},~~L_3=\{3,7,a,11,15,e,43,47\},\nonumber \\
&L_4=\{4,8,a,12,16,e,44,48\},~~M_1=\{1,5,b,17,21,f,49,53\},~~M_2=\{2,6,b,18,22,f,50,54\},\nonumber\\
&M_3=\{3,7,b,19,23,f,51,55\},~~M_4=\{4,8,b,20,24,f,52,56\},N_1=\{1,5,c,25,29,g,57,61\},\nonumber\\
&N_2=\{2,6,c,26,30,g,58,62\},~~N_3=\{3,7,c,27,31,g,59,63\},~~N_4=\{4,8,c,28,32,g,60,64\},\nonumber \\
&P_1=\{1,5,d,33,37,h,65,69\},~~P_2=\{2,6,d,34,38,h,66,70\},~~P_3=\{3,7,d,35,39,h,67,71\},\nonumber\\
&P_4=\{4,8,d,36,40,h,68,72\},\nonumber\\
&X_1=\{9,22,32,39,45,50,60,67\},~~X_2=\{10,17,27,40,46,53,63,68\},~~X_3=\{11,20,30,33,47,56,58,69\},\nonumber\\
&X_4=\{12,23,25,34,48,51,61,70\},X_5=\{13,18,28,35,41,54,64,71\},X_6=\{14,21,31,36,42,49,59,72\},\nonumber\\
&X_7=\{15,24,26,37,43,52,62,65\},X_8=\{16,19,29,38,44,55,57,66\},\nonumber\\
&Y_1=\{9,23,30,40,45,51,58,68\},~~Y_2=\{10,19,32,33,46,55,60,69\},~~Y_3=\{11,22,25,36,47,50,61,72\},\nonumber\\
&Y_4=\{12,17,26,39,48,53,62,67\},Y_5=\{13,20,27,34,41,56,63,70\},Y_6=\{14,23,28,37,42,51,64,65\},\nonumber\\
&Y_7=\{15,18,29,40,43,54,57,68\},Y_8=\{16,21,30,35,44,49,58,71\},\nonumber\\
&Z_1=\{9,24,31,38,45,52,59,66\},~~Z_2=\{10,24,25,35,46,52,61,71\},~~Z_3=\{11,17,28,38,47,53,64,66\},\nonumber\\
&Z_4=\{12,18,31,33,48,54,59,69\},Z_5=\{13,19,26,36,41,55,62,72\},Z_6=\{14,20,29,39,42,56,57,67\},\nonumber\\
&Z_7=\{15,21,32,34,43,49,60,70\},Z_8=\{16,22,27,37,44,50,63,65\}.\nonumber
\end{eqnarray}
\normalsize From there we find that $\mathcal{W}_9$ consists of 40
vertices and has spectrum $\{-4^{15},2^{24},12\}$, which are the
characteristics identical with those of the generalized quadrangle
of order three formed by the totally singular points and lines of
a parabolic quadric $Q(4,3)$ in $PG(4,3)$\cite{Payne}. The
quadrangle $Q(4,3)$, like its two-qubit counterpart, exhibits all
the three kinds of geometric hyperplanes, viz. a slim generalized
quadrangle of order (3,1) (a grid), an ovoid, and a perp-set, and
these three kinds of subsets can all indeed be found to sit inside
$\mathcal{W}_9$. One of the grids is formed by  the sixteen lines
$L_i$, $M_i$, $N_i$ and $P_i$ ($i=1$, $2$, $3$ and $4$) as
illustrated in Fig.\,6; the remaining 24 vertices comprise an
8-coclique ($X_i$, which correspond to mutually unbiased bases),
and a four-dimensional hypercube ($Y_i$ and $Z_i$). Next, one can
partition $\mathcal{W}_9$ into an independent set/coclique and the
minimum vertex cover using a standard graph software. The
cardinality of any independent set is 10 (= $3^2 + 1$), which
means that any such set is an ovoid of $Q(4,3)$\cite{Payne}. It is
easy to verify that, for example, the set $\{L_1,M_2,
N_3,P_4,X_3,X_8,Y_4,Y_6,Z_2,Z_7$\} is an ovoid; given any
ovoid/independent set, $Q(4,3)$/$\mathcal{W}_9$ can be partitioned
as shown in Fig.\,7. The remaining type of a hyperplane of
$\mathcal{W}_9$ is a perp-set, i.e. the set of 12 vertices
adjacent to a given (``reference") vertex (Fig.\,8); the set of
the remaining 27 vertices can be shown to consist of three ovoids
which share (altogether and pairwise) just a single vertex --- the
reference vertex itself. This configuration bears number $99$ in a
list of graphs with few eigenvalues given in Ref.
\citenum{Evandam} and can schematically be illustrated in form of
a \lq\lq triangle", with a triangular pattern at its nodes and a
$1 \times 2$ grid put on its edges; the union of $1 \times 2$ grid
and a triangle either forms a Mermin-square-type graph $M$, as
already encountered in the two-qubit case, or a quartic graph of
another type, denoted as $K$ (see Fig.\,8).

\begin{figure}[t]
\centerline{\includegraphics[width=7truecm,clip=]{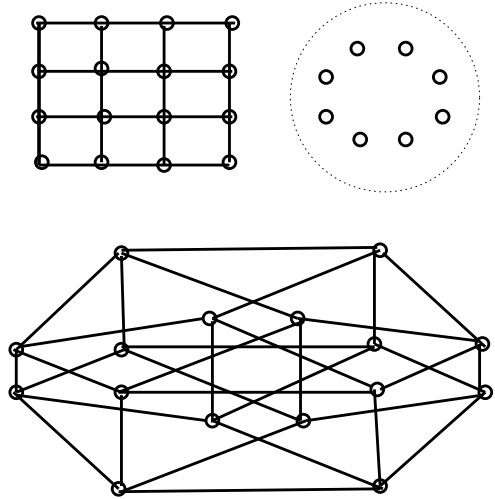}}
\caption{A partitioning of $\mathcal{W}_9$ into a grid (top left), an $8$-coclique
(top right) and a four-dimensional hypercube (bottom).}
\end{figure}
\begin{figure}[t]
\centerline{\includegraphics[width=9truecm,clip=]{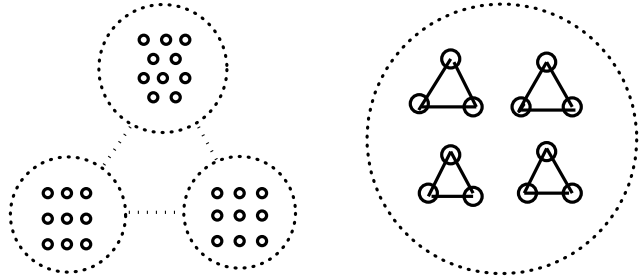}}
\caption{A partitioning of $\mathcal{W}_9$ into a tripartite graph comprising a $10$-coclique,
two $9$-cocliques and a set of four triangles; the lines corresponding to the vertices of a
selected triangle intersect at the same observables of $\mathcal{P}_9$ and the union of the
latter form a line of $\mathcal{P}_9$.}
\end{figure}

The foregoing observations and facts provide a reliable basis for us to surmise that the
geometry behind $\mathcal{W}_9$ is identical with that of $Q(4,3)$. If this is so, then
the symplectic generalized quadrangle of order three, $W(3)$, which is the dual of
$Q(4,3)$\cite{Payne}, must underlie the geometry of the Pauli graph $\mathcal{P}_9$.
However, the vertex-cardinality of $W(3)$ is 40 (the same as that of $Q(4,3)$), whilst
$\mathcal{P}_9$ features as many as 80 points/vertices. Hence,  if the geometries of
$W(3)$ and $\mathcal{P}_9$ are isomorphic, then there must exits a natural pairing between
the Pauli operators such that there exists a bijection between pairs of operators of
$\mathcal{P}_9$ and points of $W(3)$. This issue requires, obviously, a much more elaborate
analysis, to be the subject of a separate paper.

\begin{figure}[t]
\centerline{\includegraphics[width=8truecm,clip=]{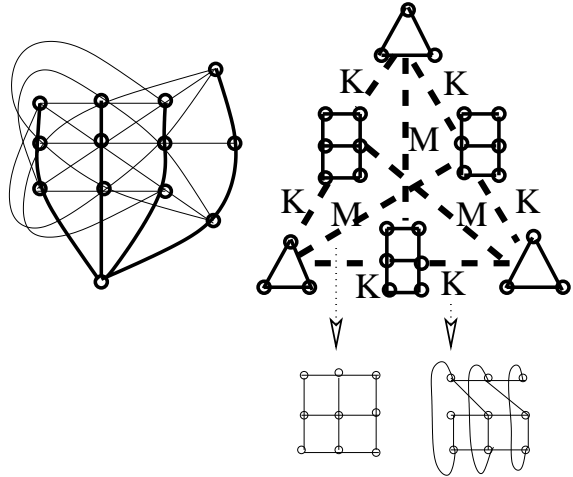}}
\caption{A partitioning of $\mathcal{W}_9$ into a perp-set and a ``single-vertex-sharing"
union of its three ovoids.}
\end{figure}

\section{$N$-qudit systems and symplectic polar spaces}
\label{polar}

We have seen that the simplest symplectic generalized quadrangles,
viz. $W(2)$ and $W(3)$, are intimately connected with the algebra
of generalized Pauli operators of, respectively, two-qubit and
two-qutrit systems. Now, it is a well-known fact that symplectic
generalized quadrangles $W(q)$, $q$ any power of a prime, are the
lowest rank symplectic polar spaces.\cite{Payne,Cameron,Ball} This
fact is enough to see how to tackle the geometry of multiple ($N$
arbitrary) qudit systems.

A symplectic polar space (see, e.\,g.,  Refs.~\citenum{Tits,Cameron,Ball} for more
details) is a $d$-dimensional vector space
over a finite field $\mathbf{F}_q$, $V(d, q)$, carrying a non-degenerate bilinear
alternating form. Such a polar space,
usually denoted as $W_{d -1}(q)$,  exists only if $d=2N$, with $N$ being its
rank. A subspace of $V(d, q)$ is called totally isotropic if
the form vanishes identically on it. $W_{2N-1}(q)$ can then be
regarded as the space of totally isotropic subspaces of $PG(2N-1,
q)$ with respect to a symplectic form, with its maximal totally isotropic subspaces,
called also generators $G$, having dimension $N - 1$.

We treat the case $q=2$, for which this
polar space contains
\begin{equation}
|W_{2N-1}(2)| = | PG(2N-1, 2)| = 2^{2N} - 1 = 4^{N} - 1
\end{equation}
points and
$ (2+1)(2^{2}+1) \ldots (2^{N}+1)$
generators. A spread $S$ of $W_{2N-1}(q)$ is a set of generators partitioning
its points.  The cardinalities of a spread and a generator of
$W_{2N-1}(2)$ are $|S| = 2^{N} + 1$ and $|G| = 2^{N} - 1$, respectively. Finally, it needs to be mentioned that two
distinct points of $W_{2N-1}(q)$ are called perpendicular if they
are joined by a line; for $q=2$, there exist $\#_{\Delta} = 2^{2N-1}$ points that are {\it not} perpendicular to
a given point.

Now, we can identify the Pauli operators of $N$-qubits
with the points of $W_{2N-1}(2)$. If, further, we identify the
operational concept ``commuting" with the geometrical one
``perpendicular," we then readily see that the points lying on
generators of  $W_{2N - 1}(2)$ correspond to maximally commuting
subsets (MCSs) of operators and a spread of $W_{2N - 1}(2)$ is
nothing but a partition of the whole set of operators into
MCSs. Finally, we get that there are $2^{2N-1}$ operators that do {\it
not} commute with a given operator.

Recognizing $W_{2N - 1}(2)$ as the geometry behind $N$-qubits, we will now turn our attention on the properties
of the associated Pauli graphs,  $\mathcal{P}_{2^N}$.
A strongly regular graph, srg$(v,D,\lambda,\mu)$, is a regular graph  having $v$ vertices and degree $D$ such that any two adjacent vertices are both adjacent to a constant number  $\lambda$ of vertices, and any two distinct non-adjacent vertices are also both adjacent to a constant number $\mu$ of vertices.  It is known that the adjacency matrix $A$ of any such graph satisfies the following equations \cite{DeClercq}
\begin{equation}
A J =D J,~~~~A^2+(\mu-\lambda)A+(\mu-D)I=\mu J,
\end{equation}
where $J$ is the all-one matrix. Hence, $A$ has $D$ as an eigenvalue with multiplicity one and its other eigenvalues are $r$ ($>0$) and $l$ ($<0$), related to each other as follows: $r+l=\lambda - \mu$ and $rl=\mu - D$. Strongly regular graphs exhibit many interesting properties \cite{DeClercq}. In particular, the two eigenvalues $r$ and $l$ are, except for (so-called) conference graphs, both integers, with
the following multiplicities
\begin{equation}
f=\frac{-D(l+1)(D-l)}{(D+rl)(r-l)}~~\mbox{and}~~g=\frac{D(r+1)(D-r)}{(D+rl)(r-l)},
\end{equation}
respectively.
The $N$-qubit Pauli graph is strongly regular, and its properties can be inferred from the relation between symplectic polar spaces and partial geometries.

A partial geometry is a more general object than a finite generalized quadrangle.  It is finite near-linear space $\{P,L\}$ such that for any point $P$ not on a line $L$, (i) the number of points of $L$ joined to $P$ by a line equals $\alpha$, (ii) each line has $(s+1)$ points, (iii) each point is on $(t+1)$ lines; this partial geometry is usually denoted as pg$(s,t,\alpha)$ \cite{Batten}.
The graph of pg$(s,t,\alpha)$ is endowed with $v=(s+1)\frac{(st+\alpha)}{\alpha}$ vertices, $\mathcal{L}=(t+1)\frac{(st+\alpha)}{\alpha}$ lines and is strongly regular of the type
\begin{equation}
{\rm srg}\left((s+1)\frac{(st+\alpha)}{\alpha},s(t+1),s-1+t(\alpha -1),\alpha(t+1)\right).
\end{equation}
The other way round, if a strongly regular graph exhibits the spectrum of a partial geometry, such a graph is called a pseudo-geometric graph.  Graphs associated with symplectic polar spaces $W_{2N-1}(q)$ are pseudo-geometric \cite{DeClercq}, being
\begin{equation}
{\rm pg}\left(q\frac{q^{N-1}-1}{q-1},q^{N-1},\frac{q^{N-1}-1}{q-1}\right)\mbox{-graphs}.
\end{equation}
Combining these facts with the findings of Sec.\,2, we conclude
that that $N$-qubit Pauli graph is of the type given by Eq.\,(11)
for $q=2$; its basics invariants for a few small values of $N$ are
listed in Table\,7.
\begin{table}[h]
\begin{center}
\begin{tabular}{|r||r|r|r||r|r|r|r||r|r|r||}
\hline
$N$& $v$ & $\mathcal{L}$ & $D$ & $r$ & $l$ & $\lambda$ & $\mu$ & $s$ & $t$ & $\alpha$\\
\hline
\hline
$2$& $15$ & $15$ & $6$ & $1$ & $-3$ & $1$ & $3$ & $2$ & $2$ & $1$ \\
\hline
$3$& $63$ & $45$ & $30$ & $3$ & $-5$ & $13$ & $15$ & $6$ & $4$ & $3$\\
\hline
$4$& $255$ & $153$ & $126$ & $7$ & $-9$ & $61$ & $63$ & $14$ & $8$ & $7$\\
\hline
\end{tabular}
\label{simpleP22}
\caption{Invariants of the Pauli graph $\mathcal{P}_{2^N}$, $N=2$, $3$ and $4$, as inferred from the properties of the symplectic polar spaces of order two and rank $N$. In general, $v=4^N -1$, $D=v-1-2^{2N-1}$, $s=2\frac{2^{N-1}-1}{2-1}$, $t=2^{N-1}$, $\alpha=\frac{2^{N-1}-1}{2-1}$, $\mu=\alpha(t+1)=rl+D$ and $\lambda=s-1+t(\alpha -1))=\mu+r+l$. The integers $v$ and $e$ can also be found from $s$, $t$ and $\alpha$ themselves.}
\end{center}
\end{table}
\section{Conclusion}
We have demonstrated that a particular kind of finite geometries, namely projective ring lines, generalized quadrangles
and symplectic polar spaces, are geometries behind finite dimensional Hilbert spaces. A detailed examination of
two-qubit (Sec.\,2) and two-qutrit (Sec.\,3) systems has revealed the fine structure of these geometries and showed
how this structure underlies the algebra of the generalized Pauli operators associated with these systems.
This study represents a crucial step towards a unified geometric picture, briefly outlined in Sec.\,4,
encompassing any finite-dimensional quantum system.

\acknowledgments     

This work was partially supported by the Science and Technology Assistance Agency under the contract $\#$
APVT--51--012704, the VEGA projects $\#$ 2/6070/26 and $\#$ 7012 (all from
Slovak Republic) and the trans-national ECO-NET project $\#$
12651NJ ``Geometries Over Finite Rings and the Properties of
Mutually Unbiased Bases" (France).

\end{document}